\documentclass{article}
\usepackage{amsmath}
\def\Ref#1{(\ref{#1})}

\begin{document}
\begin{titlepage}
\vspace*{10mm}
\begin{center}
{\large \bf On the solvable multi--species reaction--diffusion
processes} \vskip 10mm \centerline {\bf Farinaz Roshani $^a$ {\rm
and} Mohammad Khorrami $^b$} \vskip 1cm {\it Institute for
Advanced Studies in Basic Sciences,}
\\ {\it P. O. Box 159, Zanjan 45195, Iran}\\
{\it Institute for Studies in Theoretical physics and Mathematics,}\\
{\it P. O. Box 5531, Tehran 19395, Iran}\\
$^a$ roshani@iasbs.ac.ir\\
$^b$ mamwad@iasbs.ac.ir\\

\end{center}
\vskip 2cm
\begin{abstract}
\noindent A family of one--dimensional multi--species
reaction--diffusion processes on a lattice is introduced. It is
shown that these processes are exactly solvable, provided a
nonspectral matrix equation is satisfied. Some general remarks on
the solutions to this equation, and some special solutions are
given. The large--time behavior of the conditional probabilities
of such systems are also investigated.
\end{abstract}
\end{titlepage}
\newpage

\section{Introduction}
In recent years, the asymmetric exclusion process and the problems
related to it, including for example the bipolymerization
\cite{1}, dynamical models of interface growth \cite{2}, traffic
models \cite{3}, the noisy Burgers equation \cite{4}, and the
study of shocks \cite{5,6}, have been extensively investigated.
The dynamical properties of this model have been studied in
[6--8]. As the results obtained by approaches like mean field are
not reliable in one dimension, it is useful to introduce solvable
models and analytic methods to extract exact physical results.
Among these methods is the coordinate Bethe--ansatz, which was
used in \cite{Sc} to solve the asymmetric simple exclusion process
on a one--dimensional lattice. In \cite{AKK1}, a similar technique
was used to solve the drop--push model \cite{11}, and a
generalized one--parameter model interpolating between the
asymmetric simple exclusion model and the drop--push model. In
\cite{AKK2}, this family was further generalized to a family of
processes with arbitrary left- and right- diffusion rates. All of
these models were lattice models. Finally, the behaviour of latter
model on continuum was investigated in \cite{RK}. The continuum
models of this kind are also investigated in \cite{SW1,SW2}. In
\cite{RK2} a generalization of such processes was studied which
contained annihilation of particles as well.

In all of these, people have been mainly concerned with the
so--called single--species processes, in which only one kind of
particles exist and move on the lattice (or the continuum).
Another interesting problem is the study of multi--species systems
in them several kinds of particles move and interact on a lattice.
In \cite{S5}, single--species systems have been characterized for
them the equations governing the evolution of the $N$--point
functions contain $N$- or less- point functions. This has been
done for multi--species systems in \cite{AGKS}. In \cite{AA},
two--species reaction--diffusion systems have been introduced that
are solvable in the sense that the $S$ matrix corresponding to
them is factorizable into the two--particle $S$ matrices. It is
found there that the the criterion for this is that the
interactions (which are of the nearest--neighbor type) are to be
so that the $S$ matrix satisfy a kind of spectral Yang--Baxter
equation.

We follow the same line. That is, we investigate interactions
which can be written as boundary conditions for the probability
functions. These interactions preserve the total number of
particles, so that if one begins with $N$ particles, knowing the
$N$--point probabilities is enough to know everything about the
system. Using the coordinate Bethe--ansatz, it is found that for
this ansatz to be consistent, the $S$ matrix should satisfy a kind
of spectral Yang--Baxter equation \cite{AA}. However, the $S$
matrix is of a special form containing the boundary conditions (or
interactions), and not every solutions of the spectral
Yang--Baxter equation can be used to construct such a solvable
model. We investigate the spectral equation the $S$ matrix should
satisfy and show that this is equivalent to a non--spectral
equation for the boundary conditions. This is independent of the
number of species.

The scheme of the paper is the following. In section 2, the it is
shown that the prescription of investigating multi--species
reaction--diffusion systems in terms of diffusion systems equipped
with suitable boundary conditions is studied. In section 3, the
Bethe--ansatz solution for such (solvable) systems is obtained and
its large--time behavior is investigated. In section 4, the
solvability criterion is obtained and it is shown that this
criterion is a nonspectral matrix equation. In section 5, some
general properties of the solutions of the solvability criterion
is studied. Finally, in section 6 some special solutions of the
solvability equation are studied.

\section{Multi--species reaction--diffusion systems and the boundary
conditions} Consider a system consisting of $N$ particles on a
lattice, drifting to the right with unit rate if the right
neighboring site is empty, and interacting with each other only if
two of them are adjacent. Suppose that there are $n$ kinds (or
species) of particles and the interaction between the particles is
just in the form that if two particles $A^\alpha$ and $A^\beta$
are adjacent to each other, they may change to $A^\gamma$ and
$A^\delta$ with the rate $b^{\gamma\delta}_{\alpha\beta}$. That
is, the allowed processes are
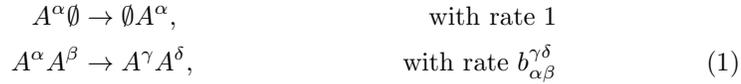
\begin{align}\label{1}
A^\alpha\emptyset&\to\emptyset A^\alpha ,\quad
&\hbox{with rate }1\nonumber\\
A^\alpha A^\beta&\to A^\gamma A^\delta ,\quad  &\hbox{with rate }
b^{\gamma\delta}_{\alpha\beta}
\end{align}
These processes result in the master equation
\begin{align}\label{2}
\dot P^{\alpha_1,\cdots ,\alpha_N}(x_1,\cdots , x_N;t)=&
     P^{\alpha_1,\cdots ,\alpha_N}(x_1-1,\cdots , x_N;t)+\cdots\nonumber\\
 &+ P^{\alpha_1,\cdots ,\alpha_N}(x_1,\cdots , x_N-1;t)\nonumber\\
 &-NP^{\alpha_1,\cdots ,\alpha_N}(x_1,\cdots , x_N;t),
\end{align}
if $x_i<x_{i+1}-1$. The symbol $P^{\alpha_1,\cdots
,\alpha_N}(x_1,\cdots , x_N;t)$ denotes the probability of finding
a particle of type $\alpha_1$ in $x_1$, a particle of type
$\alpha_2$ in $x_2$, $\cdots$ at the time $t$. The so--called
physical region consists of the points satisfying $x_i<x_{i+1}$.
If $x_i=x_{i+1}-1$, the interactions change the equation. For
clarity, let's write the evolution equation for the two--particle
sector:
\begin{align}\label{3}
\dot P^{\alpha\beta}(x,x+1)=&P^{\alpha\beta}(x-1,x+1)\nonumber\\
&+\sum_{(\gamma\delta)\ne(\alpha\beta)}b_{\gamma\delta}^{\alpha\beta}
P^{\gamma\delta}(x,x+1) -B^{\alpha\beta}\; P^{\alpha\beta}(x,x+1)
\nonumber\\
&-P^{\alpha\beta}(x,x+1),
\end{align}
where
\begin{equation}\label{3a}
B^{\alpha\beta}:=\sum_{(\gamma\delta)\ne(\alpha\beta)}
b_{\alpha\beta}^{\gamma\delta}
\end{equation}
Defining the diagonal elements of $b$ as
\begin{equation}\label{4}
b_{\alpha\beta}^{\alpha\beta}:=1-\sum_{(\gamma\delta)\ne(\alpha\beta)}
b_{\alpha\beta}^{\gamma\delta},
\end{equation}
it is seen that \Ref{3} can be written as
\begin{equation}\label{5}
\dot P^{\alpha\beta}(x,x+1)=P^{\alpha\beta}(x-1,x+1)+
b_{\gamma\delta}^{\alpha\beta}P^{\gamma\delta}(x,x+1)
-2P^{\alpha\beta}(x,x+1),
\end{equation}
where summation is implied on repeated indices. Comparing this
with \Ref{2}, it is seen that it can be written as \Ref{2}
provided one introduces the boundary condition
\begin{equation}\label{6}
P^{\alpha\beta}(x,x)=b_{\gamma\delta}^{\alpha\beta}P^{\gamma\delta}(x,x+1),
\end{equation}
or, in a more compact form,
\begin{equation}\label{7}
|P(x,x)\rangle =b\; |P(x,x+1)\rangle .
\end{equation}
The matrix $b$ should satisfy two criteria. First, its
non--diagonal elements should be nonnegative (since they are
rates). Second, the sum of the elements of each of its columns
should be one. This can be written in a compact form as
\begin{equation}\label{8}
\langle s|\otimes\langle s|b=\langle s|\otimes\langle s|,
\end{equation}
where
\begin{equation}\label{9}
s_\alpha:=1.
\end{equation}
Note that if the number of species is one, the asymmetric simple
exclusion process \cite{Sc} is obtained.

\section{The Bethe--ansatz solution}
As in \cite{AA}, one can write a Bethe--ansatz solution for
\Ref{2} with the boundary condition
\begin{equation}\label{10}
|P(\cdots,x_{k}=x,x_{k+1}=x,\cdots )\rangle =b_{k,k+1}
|P(\cdots,x_{k}=x,x_{k+1}=x+1,\cdots )\rangle,
\end{equation}
where
\begin{equation}\label{11}
b_{k,k+1}:=1\otimes\cdots\otimes
1\otimes\underbrace{b}_{k,k+1}\otimes 1 \otimes\cdots\otimes 1.
\end{equation}
We take the ansatz
\begin{equation}\label{12}
|P({\bf x};t)\rangle =e^{Et}|\Psi({\bf x})\rangle,
\end{equation}
and it is seen that $|\Psi({\bf x})\rangle$ should satisfy
\begin{align}\label{13}
E|\Psi(x_1,\cdots ,x_N)\rangle =&|\Psi(x_1-1,\cdots ,x_N)\rangle
+\cdots\nonumber\\
&+|\Psi(x_1,\cdots ,x_N-1)\rangle -N|\Psi(x_1,\cdots ,x_N)\rangle,
\end{align}
and
\begin{equation}\label{14}
|\Psi (\cdots,x_{k}=x,x_{k+1}=x,\cdots )\rangle =b_{k,k+1} |\Psi
(\cdots,x_{k}=x,x_{k+1}=x+1,\cdots )\rangle .
\end{equation}
The Bethe--ansatz is that, one takes the following form for
$|\Psi({\bf x})\rangle$.
\begin{equation}\label{15}
|\Psi({\bf x})\rangle=\sum_\sigma A_\sigma e^{i\sigma({\bf
p})\cdot {\bf x}}|\psi \rangle ,
\end{equation}
where $|\psi\rangle$ is an arbitrary vector and the summation runs
over the elements of the permutation group. Plugging this in
\Ref{13} results in
\begin{equation}\label{16}
E=\sum_{k=1}^N\left(e^{-ip_k}-1\right).
\end{equation}
The boundary condition \Ref{14} yields
\begin{equation}\label{17}
\left[ 1-e^{i\sigma(p_{k+1})}b_{k,k+1}\right] A_\sigma + \left[
1-e^{i\sigma(p_k)}b_{k,k+1}\right] A_{\sigma\sigma_k}=0.
\end{equation}
Here $\sigma$ is that element of the permutation group which only
interchanges $p_k$ and $p_{k+1}$. From this, one obtains
\begin{equation}\label{18}
A_{\sigma\sigma_k}=S_{k,k+1}[\sigma(p_k),\sigma(p_{k+1})]
A_\sigma,
\end{equation}
where the matrix $S$ is defined through
\begin{equation}\label{19}
S(p_1,p_2):=-(1-z_1\; b)^{-1}(1-z_2\; b),
\end{equation}
and the definition of $S_{k,k+1}$ is similar to that of
$b_{k,k+1}$ in \Ref{11}. we have also used the definition
\begin{equation}\label{20}
z_j:=e^{ip_j}.
\end{equation}
This shows that one can construct $A_\sigma$'s from $A_1$ by
writing $\sigma$ as a product of $\sigma_k$'s. But these elements
of the permutation group satisfy
\begin{equation}\label{21}
\sigma_k\sigma_{k+1}\sigma_k=\sigma_{k+1}\sigma_k\sigma_{k+1}.
\end{equation}
This means that
\begin{equation}\label{22}
A_{\sigma_k\sigma_{k+1}\sigma_k}=A_{\sigma_{k+1}\sigma_k\sigma_{k+1}},
\end{equation}
or
\begin{align}\label{23}
&S_{k,k+1}(p_{k+1},p_{k+2})S_{k+1,k+2}(p_k,p_{k+2})S_{k,k+1}(p_k,p_{k+1})
\nonumber\\
=&S_{k+1,k+2}(p_k,p_{k+1})S_{k,k+1}(p_k,p_{k+2})S_{k+1,k+2}(p_{k+1},p_{k+2}).
\end{align}
This can be written as
\begin{align}\label{24}
[S(p_2,p_3)\otimes 1][1\otimes S(p_1,p_3)][S(p_1,p_2)\otimes 1]=&
[1\otimes S(p_1,p_2)][S(p_1,p_3)\otimes 1]\nonumber\\
&\times[1\otimes S(p_2,p_3)].
\end{align}
Writing the $S$ matrix as the the product of the permutation
matrix $\Pi$ and an $R$ matrix:
\begin{equation}\label{25}
S_{k,k+1}=:\Pi_{k,k+1}R_{k,k+1},
\end{equation}
\Ref{24} is transformed to
\begin{equation}\label{26}
R_{23}(p_2,p_3)R_{13}(p_1,p_3)R_{12}(p_1,p_2)=R_{12}(p_1,p_2)R_{13}(p_1,p_3)
R_{23}(p_2,p_3).
\end{equation}
This is the spectral Yang--Baxter equation.

Provided this condition is satisfied, it is easy to see that the
conditional probability (the propagator) is
\begin{equation}\label{26a}
U(\mathbf{x};t|\mathbf{y};0)=\int{{d^Np}\over{(2\pi)^N}}
e^{-i\mathbf{p}\cdot\mathbf{y}}\sum_\sigma A_\sigma
e^{i\sigma(\mathbf{p})\cdot\mathbf{x}}e^{t\; E(\mathbf{p})},
\end{equation}
where the integration region for each $p_i$ is from $[0,2\pi]$,
and we have taken $A_e=1$. ($e$ is the identity of the permutation
group.) Note that \Ref{8}, and the condition of nonnegativity of
the nondiagonal elements of $b$ ensure that the absolute values of
the eigenvalues of $b$ don't exceed $1$. So there is no
singularity in $S(p_1,p_2)$ except at $p_1=0$, and this is removed
by setting $p_j\to p_j+i\epsilon$, where one should consider the
limit $\epsilon\to 0^+$. This is the same as what has been done in
\cite{Sc} and \cite{AKK1}, for example. Using this propagator, one
can of course write the probability at the time $t$ in terms of
the initial value of the probability:
\begin{equation}\label{26b}
|P(\mathbf{x};t)\rangle
=\sum_{\mathbf{y}}U(\mathbf{x};t|\mathbf{y};0)|P(\mathbf{y};0)\rangle
\end{equation}
For the two--particle sector, it is not difficult to obtain $U$.
In fact, as there is only one matrix ($b$) in the expression for
$U$, one can treat it as a $c$--number and the problem is reduced
to that of \cite{RK2}, with $\lambda$ replaced by $b$. So,
\begin{align}\label{26c}
U({\mathbf x};t|{\mathbf y};0)=&
e^{-2t}{{t^{x_1-y_1}}\over{(x_1-y_1)!}}
{{t^{x_2-y_2}}\over{(x_2-y_2)!}}\nonumber\\
&+e^{-2t}\sum_{l=0}^\infty{{t^{l+x_2-y_1}}\over{(l+x_2-y_1)!}}
{{t^{x_1-y_2}}\over{(x_1-y_2)!}}b^l \left(-1+{{t\;
b}\over{x_1-y_2+1}}\right) .
\end{align}
One can decompose the vector space on which $b$ acts into a
subspace on which $b=1$ (eigenspace of $b$ corresponding to
eigenvalue $1$) and another invariant subspace. This is done by
decomposing the unit matrix into two projectors:
\begin{equation}\label{26d}
1=Q+R,
\end{equation}
where $Q$ and $R$ are projections satisfying
\begin{equation}\label{26e}
QR=RQ=0.
\end{equation}
$Q$ is projection on the eigenspace of $b$ corresponding to the
eigenvalue $1$, and $R$ is projection on the other invariant
subspace of $b$. Using this, one can write $U$ as
\begin{align}\label{26f}
U({\mathbf x};t|{\mathbf y};0)=&
\left[e^{-2t}{{t^{x_1-y_1}}\over{(x_1-y_1)!}}
{{t^{x_2-y_2}}\over{(x_2-y_2)!}}\right.\nonumber\\
&\left.+e^{-2t}\sum_{l=0}^\infty{{t^{l+x_2-y_1}}\over{(l+x_2-y_1)!}}
{{t^{x_1-y_2}}\over{(x_1-y_2)!}}\left(-1+{t\over{x_1-y_2+1}}
\right)\right]Q\nonumber\\
&+\left[e^{-2t}{{t^{x_1-y_1}}\over{(x_1-y_1)!}}
{{t^{x_2-y_2}}\over{(x_2-y_2)!}}\right.\nonumber\\
&\left.+e^{-2t}\sum_{l=0}^\infty{{t^{l+x_2-y_1}}\over{(l+x_2-y_1)!}}
{{t^{x_1-y_2}}\over{(x_1-y_2)!}}b^l \left(-1+{{t\;
b}\over{x_1-y_2+1}}\right)\right]R .
\end{align}
Here we have used
\begin{equation}\label{26g}
b=b(Q+R)=Q+bR.
\end{equation}
As the eigenvalues of $b$, other than $1$, are assumed to have
moduli less than one, the second term in \Ref{26f} is the same as
(33) in \cite{RK2}, that is, a term obtained from the boundary
condition corresponding to annihilation ($\lambda<1$ in
\cite{RK2}). The first term corresponds to an asymmetric simple
exclusion process \cite{Sc}. The large--time behavior of these two
terms are also simply obtained. The large--time behavior of the
first was obtained in \cite{RK}, and that of the second in
\cite{RK2}. At large times, the second term is found to be
independent of $b$ (or $\lambda$) and vanishing faster than $1/t$.
Also, the summation of this term vanishes as $t$ tends to
infinity. In fact, using \cite{RK2} it is seen that
\begin{align}\label{26h}
\hbox{the second term}=&{1\over{2\pi t}}\left\{
e^{-[(x_1-y_1-t)^2+(x_2-y_2-t)^2]/(2t)}\right.\nonumber\\
&-\left. e^{-[(x_1-y_1-t)^2+(x_2-y_2-t)^2]/(2t)}\right\},\quad
t\to\infty.
\end{align}
So at large times only the first term of \Ref{26f} survives. This
means that at large times, the propagator is proportional to the
projection on the eigenspace of $b$ corresponding to the
eigenvalue 1 (the projection on the {\textit equilibrium} subspace
of $b$) and the proportionality constant is simply the propagator
of the asymmetric simple exclusion process.

To conclude, for large times the two--particle conditional
probability is that of an asymmetric simple exclusion process
projected on the eigenspace of $b$ corresponding to its unit
eigenvalue.

\section{Solvability criteria for the boundary conditions}
From \Ref{19}, it is seen that $S(p_1,p_2)$ is a binomial of
degree one with respect to $z_2:=e^{ip_2}$. Putting this in
\Ref{24}, one arrives at a quadratic expression with respect to
$z_3$. The coefficients of this expression are, of course,
matrices depending on $z_1$ and $z_2$. It is easy to find the
roots of this expression for $z_3$. In fact, putting $z_3=z_1$ in
\Ref{24}, one arrives at the identity
\begin{equation}\label{27}
[S(p_2,p_1)\otimes 1][S(p_1,p_2)\otimes 1]\equiv [1\otimes
S(p_1,p_2)][1\otimes S(p_2,p_1)].
\end{equation}
(We note that $S(p_1,p_2)S(p_2,p_1)\equiv 1$.) Also, putting
$z_3=z_2$, another identity is obtained:
\begin{equation}\label{28}
[1\otimes S(p_1,p_2)][S(p_1,p_2)\otimes 1]\equiv [1\otimes
S(p_1,p_2)][S(p_1,p_2)\otimes 1].
\end{equation}
These two identities show that the roots of the quadratic
expression for $z_3$ are $z_1$ and $z_2$. That is, one can write
that expression as
\begin{equation}\label{29}
(z_3-z_1)(z_3-z_2)Q(z_1,z_2)=0.
\end{equation}
So, \Ref{24} is equivalent to $Q=0$, which itself is obtained by
putting $z_3=0$ in \Ref{24}:
\begin{align}\label{30}
&[(1-z_2b)^{-1}\otimes 1][1\otimes(1-z_1b)^{-1}]
[(1-z_1b)^{-1}(1-z_2b)\otimes 1]\nonumber\\
=&[1\otimes(1-z_1b)^{-1}(1-z_2b)][(1-z_1b)^{-1}\otimes 1]
[1\otimes(1-z_2b)^{-1}].
\end{align}
Inverting both sides, one arrives at
\begin{align}\label{31}
&[(1-z_2b)^{-1}(1-z_1b)\otimes
1][1\otimes(1-z_1b)][(1-z_2b)\otimes 1] \nonumber\\
=&[1\otimes(1-z_2b)][(1-z_1b)\otimes
1][1\otimes(1-z_2b)^{-1}(1-z_1b)].
\end{align}
This is a quadratic expression in terms of $z_1$. For $z_1=0$,
\Ref{31} gives the identity
\begin{equation}\label{32}
[(1-z_2b)^{-1}\otimes 1][(1-z_2b)\otimes 1]\equiv
[1\otimes(1-z_2b)] [1\otimes(1-z_2b)^{-1}],
\end{equation}
while for $z_1=z_2$, the identity
\begin{equation}\label{33}
[1\otimes(1-z_2b)][(1-z_2b)\otimes 1]\equiv [1\otimes(1-z_2b)]
[(1-z_2b)\otimes 1]
\end{equation}
is obtained. So, the quadratic expression corresponding to
\Ref{31} is equivalent to
\begin{equation}\label{34}
z_1(z_1-z_2)\tilde Q(z_2)=0,
\end{equation}
and to find $\tilde Q$, one simply uses the coefficient of $z_1^2$
in \Ref{31}. This is
\begin{equation}\label{35}
(1-z_2b_{12})^{-1}b_{12}b_{23}(1-z_2b_{12})=(1-z_2b_{23})b_{12}b_{23}
(1-z_2b_{23})^{-1},
\end{equation}
or
\begin{equation}\label{36}
b_{12}b_{23}(1-z_2b_{12})(1-z_2b_{23})=(1-z_2b_{12})(1-z_2b_{23})b_{12}
b_{23}.
\end{equation}
This is a quadratic expression in $z_2$. But the coefficients of
$z_2^0$ and $z_2^2$ are identities. So the only remaining equation
is
\begin{equation}\label{37}
b_{12}b_{23}(b_{12}+b_{23})=(b_{12}+b_{23})b_{12}b_{23},
\end{equation}
or
\begin{equation}\label{38}
b_{12}[b_{12},b_{23}]=[b_{12},b_{23}]b_{23}.
\end{equation}
\Ref{38} is equivalent to \Ref{24}. But it is seen that \Ref{38}
is non--spectral, whereas \Ref{24} is spectral. So it is far
simpler to seek the solutions to \Ref{38} than to seek those of
\Ref{24}.

To summarize, a matrix $b$, or the reactions \Ref{1}, correspond
to an exactly solvable reaction diffusion system on a
one--dimensional lattice, provided $b$ satisfies \Ref{38} and
\Ref{4} (or \Ref{8}, equivalently), and the non--diagonal elements
of $b$ are nonnegative.

\section{General properties of the solutions to the solvability
criteria} Solutions to \Ref{38} enjoy two general properties.
First, if $b$ is a solution, then
\begin{equation}\label{40}
b':=\alpha b+\beta,
\end{equation}
is another solution for constant $\alpha$ and $\beta$. If $b$
satisfies \ref{8}, then
\begin{equation}\label{41}
\langle s|\otimes\langle s|b'=(\alpha +\beta)\langle
s|\otimes\langle s|.
\end{equation}
So, putting $\beta :=1-\alpha$, ensures that $b'$ satisfies
\Ref{8}. If $\alpha>0$, then the nondiagonal elements of $b'$ are
nonnegative provided the nondiagonal elements of $b$ are
nonnegative. So
\begin{equation}\label{42}
b':=\alpha b+(1-\alpha )
\end{equation}
corresponds to a solvable system (for $\alpha>0$) if $b$ does. It
is easy to see that the meaning of this transformation is simply
to multiply the reaction rates by $\alpha$.

Second, if $b$ is a solution to \Ref{38}, then
\begin{equation}\label{43}
b':=u\otimes u\; b\; u^{-1}\otimes u^{-1}
\end{equation}
satisfies \Ref{38} as well. Here $u$ is an arbitrary (nonsingular)
matrix. This transformation, however, does not necessarily
respects the conditions \Ref{8} and nonnegativity of the rates.
So, another problem arises. Suppose $b$ is a solution to \Ref{38},
and we want to obtain a solvable system using the transformation
\Ref{43}. We must have
\begin{equation}\label{44}
\langle s|\otimes\langle s|u\otimes u\; b=\langle s|\otimes\langle
s|u\otimes u.
\end{equation}
This means that $u$ must change $\langle s|$ to some $\langle s'|$
so that $\langle s'|\otimes\langle s'|$ is a left eigenvector of
$b$ corresponding to a unit eigenvalue. One may search in the
eigenvectors of $b$ to find whether there is an eigenvector of the
form $\langle s'|\otimes\langle s'|$. If there is such an
eigenvector, then any matrix $u$ which changes $\langle s|$ to
$\langle s'|$ can be used to obtain $b'$ according to \Ref{43}.
This $b'$ satisfies \Ref{38} and \Ref{8}. But its diagonal
elements may be nonnegative or not; this should be checked
separately. If non of the eigenvalues of $b$ are of the form
$\langle s'|\otimes\langle s'|$, then this method cannot be used
to obtain a solvable system. This method resembles very much to
that used in \cite{AK}.

\section{some special cases}
{\bf case I}: $b^2=\alpha +\beta b$ ($\alpha$ and $\beta$ are
numbers). In this case, one can define
\begin{equation}\label{45}
b':=b+\gamma,
\end{equation}
with $\gamma$ satisfying
\begin{equation}\label{46}
\gamma^2+\beta\gamma-\alpha=0,
\end{equation}
to obtain
\begin{equation}\label{47}
b^{\prime 2}=(\beta +2\gamma)b'.
\end{equation}
Putting this $b'$ in \Ref{38}, one obtains the braid equation for
$b'$:
\begin{equation}\label{48}
b'_{12}b'_{23}b'_{12}=b'_{23}b'_{12}b'_{23}.
\end{equation}
From \Ref{47}, it is seen that $b'$ either can be scaled to a
projection ($b^{\prime 2}=b'$), or is nilpotent. One concludes
then that any nilpotent or projection solution to the
(nonspectral) braid equation is a solution to \Ref{38}. One can
then use any linear combination of this solution with the unit
matrix as another solution to that equation. Note, however, that
these solutions of \Ref{38} do not necessarily satisfy other
criteria of the solvable system, that is nonnegativity of the
nondiagonal elements and \Ref{8}. An inspection of the solutions
obtained in \cite{AA} shows that solutions 1---15, and 17 are of
this type. As mentioned in the previous section, one can of course
take a linear combination of each solution with the unit matrix to
obtain another solution.

{\bf case II}: $b=u\otimes v$. Here \ref{38} takes the form
\begin{equation}\label{49}
u^2\otimes v[v,u]\otimes v=u\otimes [v,u]u\otimes v^2.
\end{equation}
A simple way to satisfy this is to set
\begin{equation}\label{50}
[u,v]=0.
\end{equation}
So, using any two commuting matrices $u$ and $v$ one can construct
a solution to \Ref{38}. If the elements of one of these matrices
are nonnegative, and the nondiagonal elements of the other are
also nonnegative, then the nondiagonal elements of $b$ are
nonnegative. If
\begin{equation}\label{51}
\langle s|u=\langle s|v=\langle s|,
\end{equation}
then $b$ satisfies \Ref{8} as well. Of course, having found a
solution of this type one can use a linear combination of it with
the unit matrix as another solution. Solutions 1, 4, 7, 14, 17,
20, 21, 22, 25, 26, and 28 of \cite{AA} are of this kind.

It is possible to have other solutions to \Ref{49}. In this case,
let's also use \Ref{8}. This shows that one may rescale $u$ and
$v$ so that \Ref{51} is satisfied. One then arrives at
\begin{align}\label{52}
u^2&=u\nonumber\\
v^2&=v\nonumber\\
vuv&=uvu,
\end{align}
if $[u,v]\ne 0$. From these, it is seen that $u$, $v$, $1-u$, and
$uvu$ are projections. Moreover,
\begin{equation}\label{53}
(1-u)uvu=uvu(1-u)=0.
\end{equation}
This shows that $1-u$ and $uvu$ can be simultaneously
diagonalized. The diagonal form of them will be
\begin{align}\label{54}
1-u&=
  \begin{pmatrix}
    0&0&0\\
    0&0&0\\
    0&0&1\\
  \end{pmatrix}\nonumber\\
uvu&=
  \begin{pmatrix}
    1&0&0\\
    0&0&0\\
    0&0&0\\
  \end{pmatrix}
\end{align}
Here the elements of the above matrices are matrices themselves,
and 1 is the unit matrix of the appropriate dimension. Writing an
ansatz for $v$:
\begin{equation}\label{55}
v=\begin{pmatrix}
    v_{11}&v_{12}&v_{13}\\
    v_{21}&v_{22}&v_{23}\\
    v_{31}&v_{32}&v_{33}\\
  \end{pmatrix}
\end{equation}
and putting it in \Ref{52}, one finally arrives at the following
forms for $u$ and $v$.
\begin{align}\label{56}
u&=\begin{pmatrix}
    1&0&0&0\\
    0&1&0&0\\
    0&0&0&0\\
    0&0&0&0\\
  \end{pmatrix}\nonumber\\
v&=\begin{pmatrix}
    1&0&0&0\\
    0&0&w&0\\
    0&w'&1&0\\
    0&0&0&0\\
  \end{pmatrix}
\end{align}
where all of the entries in the above matrices are matrices, and
$w$ and $w'$ should satisfy
\begin{equation}\label{57}
ww'=w'w=0.
\end{equation}
Each of the diagonal blocks of these matrices may be zero
dimensional, except the first. It should, at least, be one
dimensional. The reason is that $u$ and $v$ have at least one
common left eigenvector, $\langle s|$, corresponding to the unit
eigenvalue. Also the dimension of each block of $u$ is equal to
that of the corresponding block in $v$. Also note that if the
dimension of $u$ and $v$ is 2 (there are two kinds of particles)
then ther will no space left for $w$ and $w'$, and $u$ and $v$
must be commuting.

The final result is that in two dimension no new solution exists
($u$ and $v$ must be commuting), and in more than two dimensions,
$u$ and $v$ must be of the form \Ref{56}. Of course any similarity
transformation on \Ref{56} gives another solution to \Ref{38}. In
fact, one has to use a similarity transformation to make $\langle
s|$ a left eigenvector of $u$ and $v$ with unit eigenvalue.

Two very simple subcases are $b=1\otimes v$ and $b=u\otimes 1$.
These describe reactions
\begin{equation}\label{58}
A^\alpha A^\beta\to A^\alpha A^\delta ,\quad\hbox{with rate }
v^\delta_\beta,
\end{equation}
and
\begin{equation}\label{59}
A^\alpha A^\beta\to A^\gamma A^\beta ,\quad\hbox{with rate }
u^\gamma_\alpha,
\end{equation}
respectively. That is, in each case only one of the particles
change, and the rate of change is independent of the type of the
other particle.


\newpage
\begin{thebibliography}{99}
\bibitem{1}   C. T. MacDonald, J. H. Gibbs, \& A. C. Pipkin; Biopolymers
              {\bf 6} (1968) 1.
\bibitem{2}   J. Krug \& H. Spohn; in {\it Solids far from equilibrium},
              edited by C. Godreche (Cambridge University Press, Cambridge,
              England, 1991), and references therein.
\bibitem{3}   K. Nagel; Phys. Rev. {\bf E53} (1996) 4655.
\bibitem{4}   J. M. Burgers, {\it The nonlinear diffusion equation} (Reidel,
              Boston, 1974).
\bibitem{5}   B. Derrida, S. A. Janowsky, J. L. Lebowitz, \& E. R. Speer;
              Europhys. Lett. {\bf 22} (1993) 651.
\bibitem{6}   P. A. Ferrari \& L. R. G. Fontes; Probab. Theory Relat. Fields
              {\bf 99} (1994) 305.
\bibitem{7}   T. Ligget, {\it Interacting particle systems} (Springer Verlag,
              New York, 1985).
\bibitem{8}   L. H. Gava \& H. Spohn; Phys. Rev. {\bf A46} (1992) 844.
\bibitem{Sc}  G. M. Sch\"utz; J. Stat. Phys. {\bf 88} (1997) 427.
\bibitem{AKK1}M. Alimohammadi, V. Karimipour, \& M. Khorrami; Phys. Rev.
              {\bf E57} (1998) 6370.
\bibitem{11}  G. M. Sch\"utz \& E. Domnay; J. Stat. Phys. {\bf 72} (1993)
              277.
\bibitem{AKK2}M. Alimohammadi, V. Karimipour, \& M. Khorrami; J. Stat.
              Phys. {\bf 97} (1999) 373.
\bibitem{RK}  F. Roshani \& M. Khorrami; Phys. Rev. {\bf E60} (1999) 3393.
\bibitem{SW1} T. Sasamoto \& M. Wadati; J. Phys. Soc. Japan {\bf E67} (1998)
              784.
\bibitem{SW2} T. Sasamoto \& M. Wadati; J. Phys. {\bf A31} (1998) 6057.
\bibitem{RK2} F. Roshani \& M. Khorrami; cond--mat/0007352.
\bibitem{S5}  G. M. Sch\"utz; J. Stat. Phys. {\bf 79} (1995) 243.
\bibitem{AGKS}A. Aghamohammadi, A. H. Fatollahi, M. Khorrami, \&
              A. Shariati; Phys. Rev. {\bf E62} (2000) 4642.
\bibitem{AA}  M. Alimohammadi \& N. Ahmadi; Phys. Rev. {\bf E62}
              (2000) 1674.
\bibitem{AK}  A. Aghamohammadi \& M. Khorrami; J. Phys. {\bf A33} (2000) 7843.

\end{thebibliography}
\end{document}